\documentclass[12pt,preprint]{aastex}

\newcommand{\bm}[1]{\mbox{\boldmath{$#1$}}}

\shorttitle{Turbulence in Galaxy Clusters}
\shortauthors{Fujita}

\begin{document}

\title{A Simple Measurement of Turbulence in Cores of Galaxy Clusters}

\author{Yutaka Fujita} 
\affil{National Astronomical Observatory, Osawa
2-21-1, Mitaka, Tokyo 181-8588, Japan, and Department of Astronomical
Science, The Graduate University for Advanced Studies, Osawa 2-21-1,
Mitaka, Tokyo 181-8588, Japan.}
\email{yfujita@th.nao.ac.jp}

\begin{abstract}
 Using a simple model, we study the effects of turbulence on the motion
 of bubbles produced by AGN jet activities in the core of a galaxy
 cluster. We focus on the turbulence with scales larger then the size
 of the bubbles. We show that for a bubble pair with an age of $\sim
 10^8$~yr, the projected angle between the two vectors from the cluster
 center to the two bubbles should be $\gtrsim 90^\circ$ and the ratio of
 their projected distances from the cluster center should be $\lesssim
 2.5$, if the velocity and scale of the turbulence are $\sim 250\rm \:
 km\: s^{-1}$ and $\sim 20$~kpc, respectively. The positions of the
 bubbles observed in the Perseus cluster suggest that the turbulent
 velocity is $\gtrsim 100\:\rm km\: s^{-1}$ for the cluster.
\end{abstract}


\section{Introduction}

In the central region of a galaxy cluster, the radiative cooling time of
intracluster medium (ICM) is generally much smaller than the Hubble
time. In the absence of any heating sources, this means that the ICM
flows subsonically toward the cluster center with a mass deposition rate
of $\dot{M}\sim 100$--$1000\: M_\sun$ \citep{fab94}. This flow had been
known as a `cooling flow'.  However, {\it ASCA} and {\it XMM-Newton}
failed to detect emission from low temperature gas, implying that the
actual cooling rate is at least 5 or 10 times smaller than that
previously assumed \citep[e.g.][]{ike97,mak01,pet01,tam01,kaa01}. These
observations strongly suggest that the mass deposition must be prevented
by some additional source of heat that balances radiative losses.

Turbulence in the ICM may work as a heating source. Through turbulent
mixing, it transfers thermal energy from the outer layer of a cluster
into the core \citep{cho03,kim03,voi04}. Dissipation of the turbulence
can also heat the core. The evidence of turbulence has indirectly
observed.  For example, \citet{chu04} found that there is little
evidence of resonant scattering in the Perseus cluster, which suggests
motion of the ICM. Recently, \citet*{fuj04} showed that bulk gas motion
induced by cluster mergers and radiative cooling develop turbulence in
and around a core. The reason is that the cool and dense core cannot be
moved by the bulk gas motion, and the resultant relative gas motion
between the core and the surrounding ICM leads to hydrodynamic
instabilities. The core is heated by the turbulence. Since the
turbulence is produced and the heating is effective only when the core
is cooling and dense, fine-tuning of balance between cooling and heating
is alleviated for this model.  On the other hand, the activities of the
active galactic nuclei (AGNs) at cluster centers may also create
turbulence.  In particular, the bubbles formed through the AGN
activities may move outward in the ICM via buoyancy and may create
turbulence behind them \citep*{chu01,qui01,sax01}.

These two types of turbulence have different features. The typical scale
of the turbulence produced by the bulk gas motion in the ICM is close to
the size of a cluster core \citep[$\gtrsim
20$~kpc;][]{fuj04,fuj05}. This scale is larger than the typical size of
a bubble \citep[$\sim 10$~kpc;][]{fab03a}.  Thus, the bubble, if any,
will be staggered in the velocity fields of the turbulence. On the other
hand, the maximum scale of the turbulence produced by a buoyant bubble
should be limited to the size of the bubble. Since the bubble is the
engine of the turbulence, its motion should not be affected much by the
turbulence.

Recently, old bubbles have been observed in the X-ray and radio
bands. They are located away from cluster centers. In the Virgo cluster,
diffuse radio emission from possible old bubbles was observed at a
distance of $\gtrsim 20$~kpc from the cluster center \citep{owe00}. From
X-ray observations, `arms' of cold gas have been found; they might be
cold gas entrained by the old bubbles from the cluster center
\citep{chu01,bel01,mol02,you02}. The arms may indicate the paths of the
bubbles. For the Perseus cluster, two bubble pairs (four bubbles) have
been found by {\it Chandra} \citep[][]{fab02}. While the inner pair is
at $r\sim 5$~kpc from the cluster center, the outer pair is at $r\sim
40$~kpc.

If a bubble pair is formed through AGN jet activities, the initial
positions of the two bubbles are expected to be symmetric about the
cluster center, where the AGN is located. There are several clusters for
which high-resolution X-ray and radio observations were made and the
positions of newborn twin bubbles were obtained. Among such clusters,
Hydra~A, A2052, M84, Cygnus~A, and Centaurus respectively have two
bubbles that are next to the central AGNs and the positions are
consistent with being symmetric about the AGNs
\citep*{mcn00,bla01,fin01,smi02,fab05}. On the other hand, the positions
of the two inner bubbles in the Perseus cluster do not appear to be
symmetric about the AGN \citep[the upper panel of Fig.~1
in][]{fab02}. However, radio observations showed that the directions of
jets are symmetric and a new bubble pair appears to be forming in those
directions \citep[the upper panel of Fig.~3 in][]{fab02}. The
complicated shapes of the inner bubbles may be due to the projection of
a few bubble pairs.

In contrast with the newborn bubbles, the positions of the old bubbles
are not always symmetric about the cluster centers.  For the Virgo
cluster, the angle between the two arms is $\sim 135^\circ$ (Figure~1
[upper right] of \citealt{chu01}). Moreover, the arms are not straight
and they are bent at $r\sim 20$~kpc \citep{for05}. For the Perseus
cluster, although no clear arms have been observed, the angle between
the two vectors from the cluster center to the two outer old bubbles is
$\sim 120^\circ$. In this letter, we study how the large-scale
turbulence, which is not created by bubble motions, affects the motion
of the bubbles and makes the observed asymmetries.  We use the Hubble
constant of $H_0=70\:\rm km\: s^{-1}\: Mpc^{-1}$.

\section{Models}

We assume that the gravitational potential of a cluster is spherically
symmetric and the central AGN creates two spherical bubbles near the
cluster center. At $t=0$, the bubbles are at a radius $R_0$ from the
cluster center in the opposite directions with respect to the cluster
center. Each bubble moves outward in the cluster via buoyancy. The
equation of motion for a bubble is
\begin{equation}
 m_{\rm bub}\frac{d\bm{u}}{dt}=m_{\rm bub}\bm{g}
-\rho |\bm{u}-\bm{u}_t|(\bm{u}-\bm{u}_t) \pi r_{\rm bub}^2
-\frac{4}{3} \pi r_{\rm bub}^3
\nabla P
\:,
\end{equation}
where $m_{\rm bub}$ is the mass of the bubble, $\bm{u}$ is the velocity
of the bubble, $\bm{u}_t$ is the velocity of turbulence at the position
of the bubble, $\bm{g}$ is the gravitational acceleration, $r_{\rm bub}$
is the radius of the bubble, and $P$ is the ICM pressure at the position
of the bubble. We assume that a bubble is adiabatic and in pressure
equilibrium with the surrounding ICM. Thus, the radius of the bubble has
the relation of
\begin{equation}
 r_{\rm bub}=r_{\rm bub, 0}(P/P_0)^{1/3\gamma}\:,
\end{equation}
where $r_{\rm bub, 0}$ is the initial radius of the bubble, $P_0$ is
the ICM pressure at the position where the bubble is created, and
$\gamma$ ($=4/3$) is the adiabatic constant.

For the sake of simplicity, we assume that the turbulence has a
monochromatic spectrum with a velocity of $u_t$ and a spatial scale of
$l_t$. We divide a cluster into small cubes with a side of $l_t$. We
assume that in each cube, the velocity field of the ICM is uniform, that
is, the velocity and the direction are constant. The direction has no
correlation among the cubes.

For parameters of a model cluster, we use the results of {\it
XMM-Newton} observations of the Perseus cluster. We chose the cluster
because there were a few studies about turbulence in the cluster
\citep{chu04,reb05}. \citet{chu03} obtained the density and temperature
profiles of the cluster (their equations~[4] and~[5]). We assume that
the ICM is in pressure equilibrium, because the velocity of the
turbulence is smaller than the sound velocity of the ICM. From the
density and temperature profiles, ignoring the self-gravity of the ICM,
we derived the mass profile of the cluster, $M(R)$, where $R$ is the
distance from the cluster center.

\section{Results and Discussion}

We assume that the initial radii of bubbles are $r_{\rm bub, 0}=10$~kpc
and their initial positions are $R_0=10$~kpc. The bubble mass, $m_{\rm
bub}$, is 0.01 times the mass of the ambient gas with the same
volume. The results are not sensitive to $m_{\rm bub}$, as long as it is
small. We calculated the motion of bubbles for several kinds of
turbulence. Since $r_{\rm bub}\sim 10$~kpc, we limited $l_t$ to be $\geq
10$~kpc. The model parameters are shown in Table~\ref{tab:model}. They
are based on the results of numerical simulations of a cluster core
\citep*{fuj04,fuj05}.  Other studies also suggested the existence of
similar velocity fields in cluster cores. \citet{den05} studied the
balance between turbulent heating and radiative cooling in
clusters. Comparing their model with observations, they concluded that
$100\lesssim u_t\lesssim 300\:\rm km\: s^{-1}$ if $l_t=\alpha R+l_0$,
where $0.05<\alpha<1$ and $l_0=0.5$~kpc. \citet{reb05} also obtained the
values similar to those shown in Table~1 from the observed metal
distribution in the core of the Perseus cluster. Note that while the
method invented by \citet{reb05} measures the velocity fields averaged
on a time-scale of $\gtrsim$~Gyr (the time-scale of metal ejection from
stars), our method measures those averaged on a time-scale of $\sim
0.1$~Gyr (the age of bubbles; see below).

Since the expansion velocity of a bubble at the formation is larger than
$u_t$ in Table~\ref{tab:model} \citep*[e.g.][]{sok02}, we assume that
the initial velocity of the bubbles is $u=0$.  For each turbulence
model, we randomly chose the initial direction of the line connecting
the two bubbles, the position of the cluster center in the cube that
contains the cluster center, and the viewing angle; they are independent
each other. For each model, we calculated the motions of 10,000 bubble
pairs and their projected positions at $t=10^8$~yr.

\begin{figure}
\epsscale{0.45} \plotone{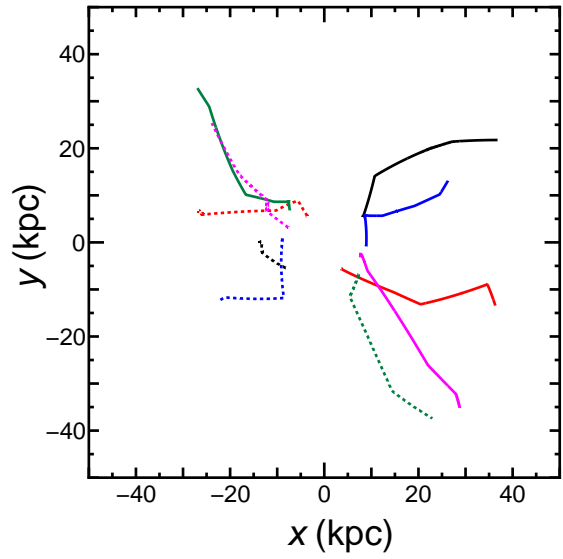} \caption{Projected trajectories for
five randomly selected bubble pairs. A pair is represented by the solid
and dotted lines with the same color. The cluster center is
 $(x,y)=(0,0)$ \label{fig:path}}
\end{figure}

Figure~\ref{fig:path} shows the projected trajectories of five randomly
selected bubble pairs for $l_t=20$~kpc and $u_t=250\:\rm km\: s^{-1}$
(Model~A2). The end points correspond to $t=10^8$~yr. We found that the
bubbles reach projected distances of $b\sim 10$--50~kpc from the cluster
center, which are similar to the positions of the outer bubbles observed
in the Perseus cluster (33 and~38~kpc). Although there are exceptions,
the bubbles move fairly straight from their departure points (at least
not zigzag), because $u_t$ is smaller than the velocity driven by
buoyancy alone ($\sim 440\rm\: km\: s^{-1}$).  This is almost the same
for other models except for model~B2 ($u_t=500\:\rm km\: s^{-1}$).

\begin{figure}
\epsscale{1.0} 
\plottwo{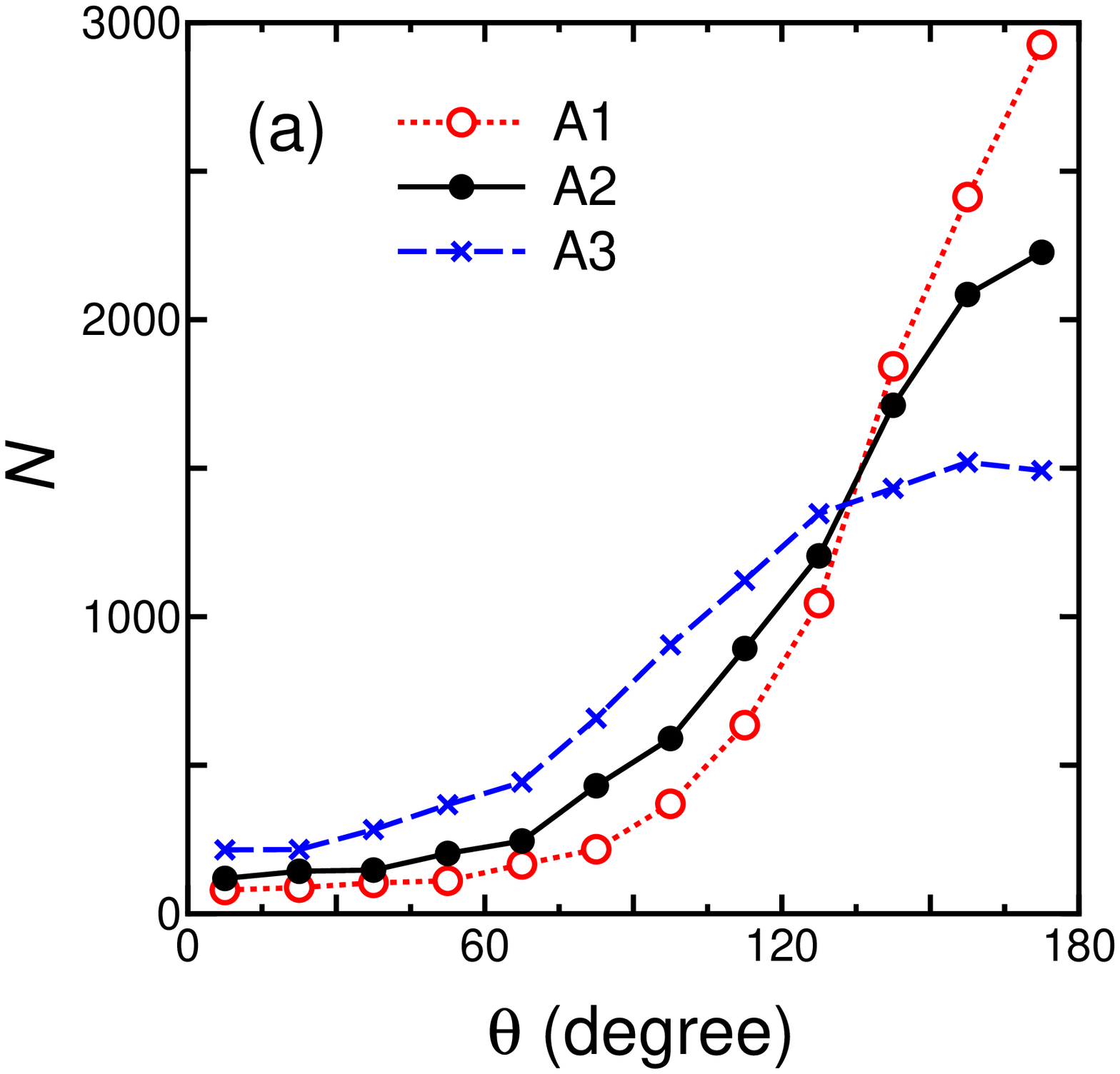}{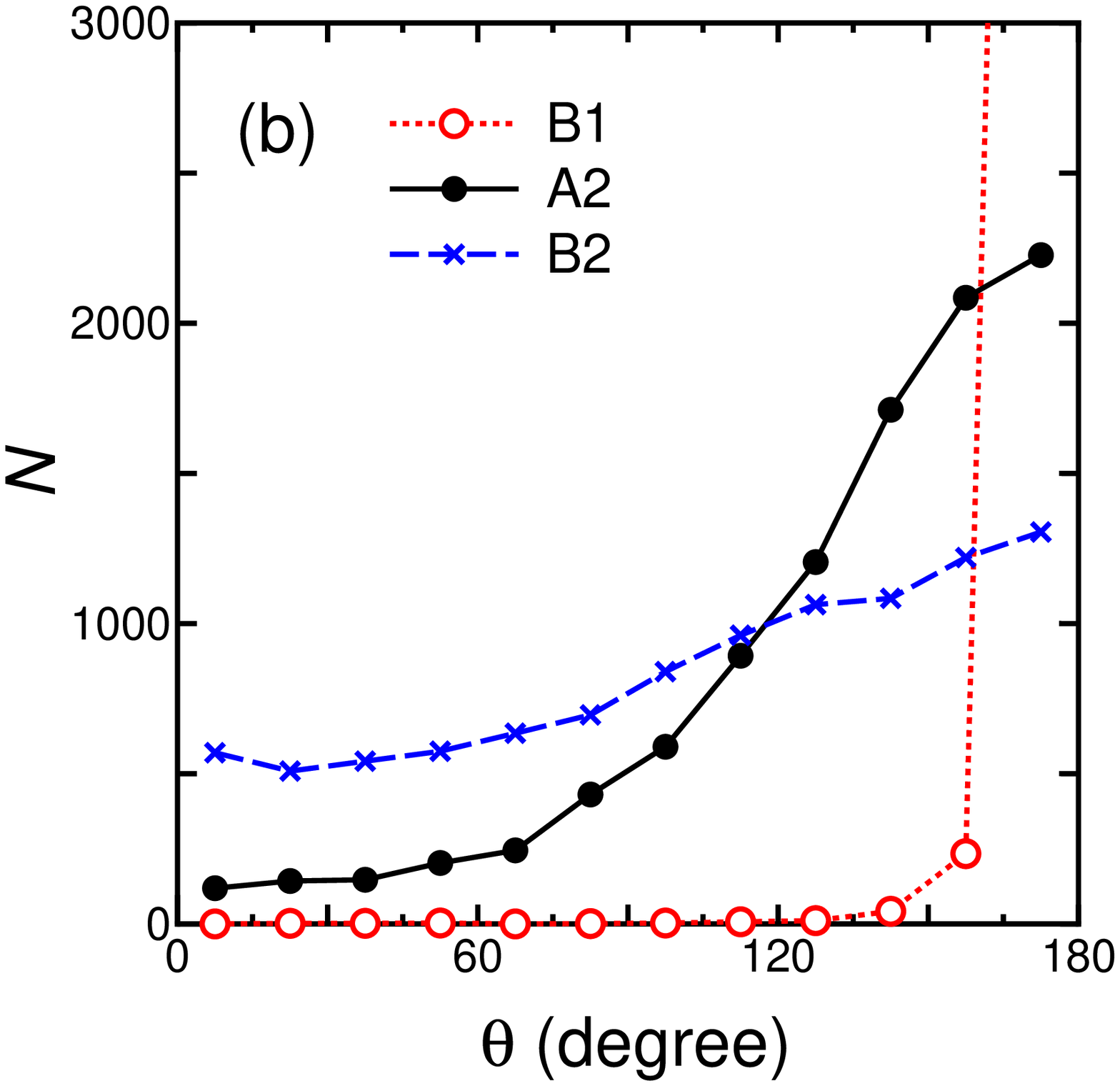} \caption{(a) Histogram of the angle $\theta$
 for models~A1--A3. (b) Same as (a) but for models~B1 and B2.
\label{fig:angle}}
\end{figure}

Figure~\ref{fig:angle} shows the distributions of the projected angle
$\theta$ between the two vectors from the cluster center to the two
bubbles composing a pair. If the two bubbles are located in the opposite
directions with respect to the cluster center, $\theta=180^{\circ}$. In
model~A2, for example, about 90\% of bubble pairs have $\theta \gtrsim
90^\circ$. The angle $\theta$ has a wider distribution for larger $u_t$
and $l_t$ because the bubbles are scattered more effectively by the
turbulence. This scatter works as diffusion in the
$\theta$-space. Medians of the angle are $\theta_m\sim
120$--$150^{\circ}$ except for model~B1 (Table~\ref{tab:model}). In
model~B1, almost all bubble pairs have $\theta\approx 180^{\circ}$,
because $u_t$ is much smaller than the velocity driven by buoyancy. In
model~B2, there is no preferential angle (Figure~\ref{fig:angle}b). For
the Perseus cluster, the angle between the two outer bubbles is
$\theta=120^\circ$. In Table~\ref{tab:model}, we show the fraction of
bubble pairs with $\theta>120^\circ$. The values of
$f(\theta>120^\circ)$ indicate that the angle distributions of all the
models except for~model~B1 are consistent with the observation.  In
model~B1, the value of $\theta$ for the Perseus cluster is unacceptably
small (the probability is $<$1\%), which means that $u_t$ is too small
to account for the observation. For the cluster, $u_t\gtrsim 100\rm\:
km\: s^{-1}$ is required, which could be detected by {\it SUZAKU}.

\begin{figure}
\epsscale{1.0} 
\plottwo{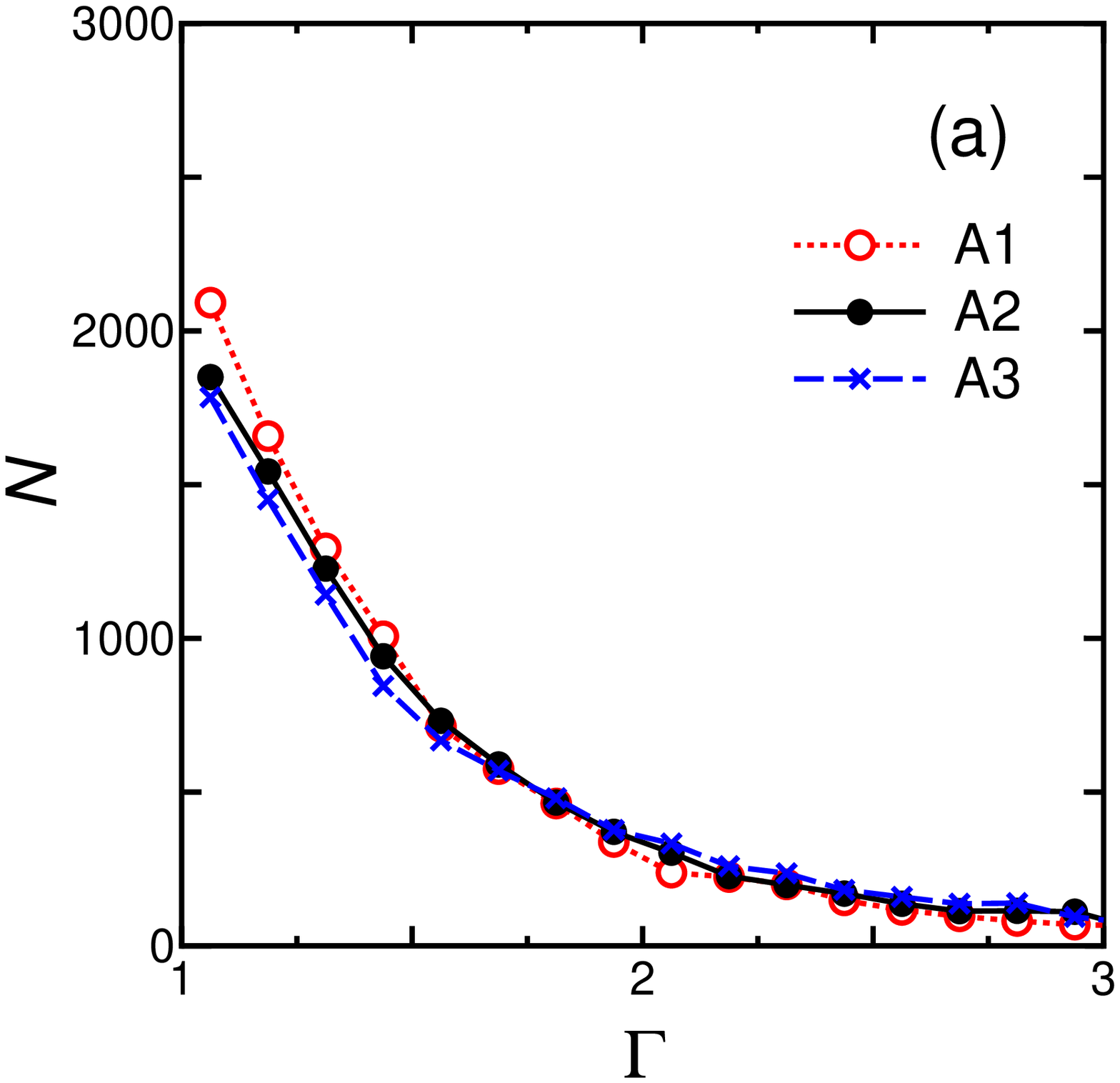}{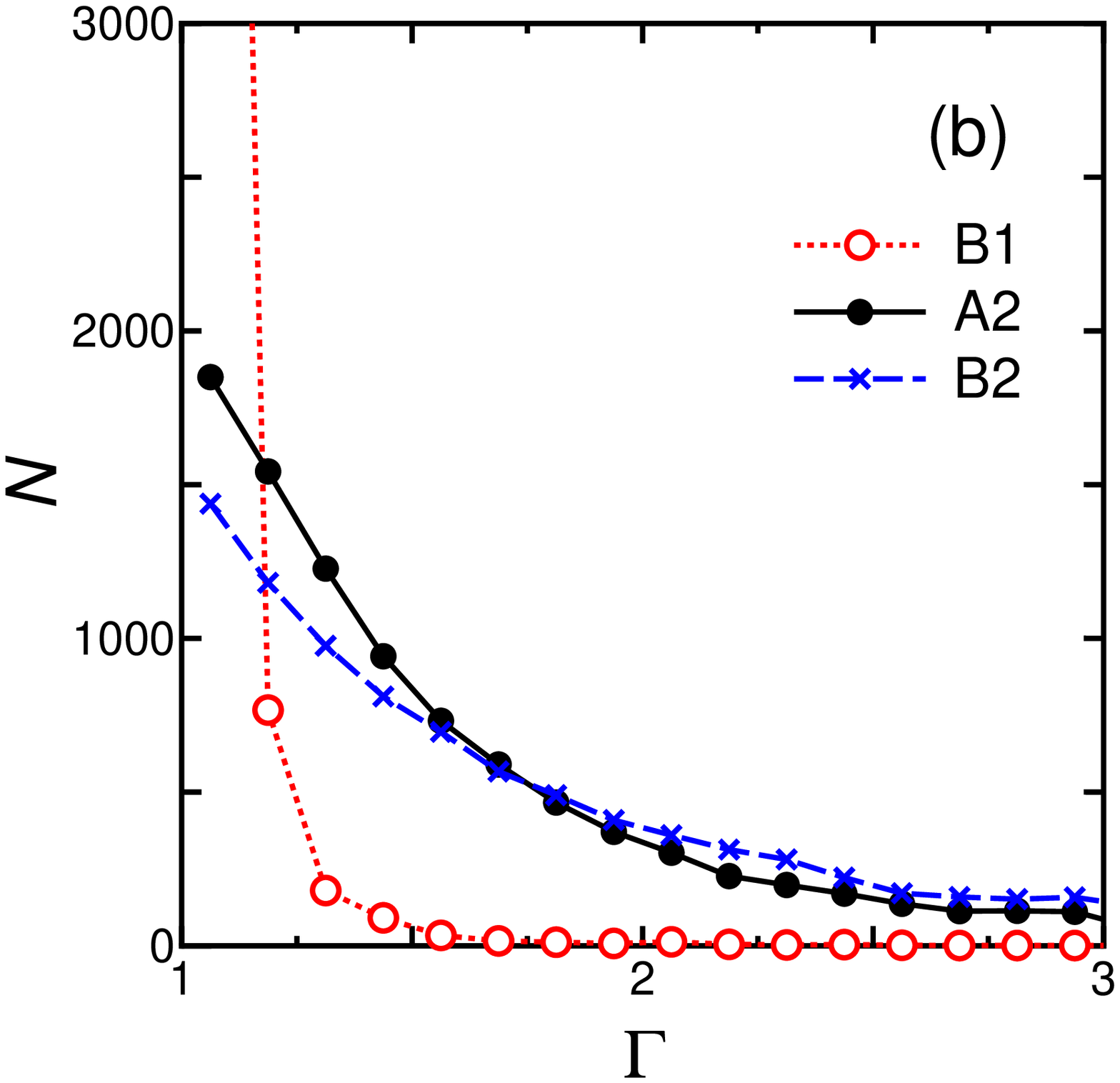} \caption{(a) Histogram of the ratio $\Gamma$
 for models~A1--A3. (b) Same as (a) but for models~B1 and B2.
\label{fig:ratio}}
\end{figure}

We refer to $b_i$ ($i=1,2$) as the projected distances of the two
bubbles composing a pair from the cluster center.
Figure~\ref{fig:ratio} shows the distributions of the radio
$\Gamma=\max(b_1/b_2, b_2/b_1)$. In model~A2, for example, about 90\%
of bubble pairs have $\Gamma \lesssim 2.5$. As is the case of $\theta$,
the ratio has a wider distribution for larger $u_t$ and $l_t$. Medians
of the ratio, $\Gamma_m$, are also shown in Table~\ref{tab:model}. For
the two outer bubbles in the Perseus cluster, the ratio is 1.15, and the
fractions of bubble pairs with a ratio larger than this,
$f_\Gamma(>1.15)$, are shown in Table~\ref{tab:model}. All models are
consistent with the observation, although model~B1 is less
preferable. We also studied the two dimensional distributions of
$\theta$ and $\Gamma$. We found no strong correlation between them.

Since the effective diffusion coefficient for the bubble motion is
represented by $D\propto u_t l_t$, models~A3 and~B2 have the same
coefficient. However, the angle and ratio distributions of the two
models are different (Figures~\ref{fig:angle} and~\ref{fig:ratio});
those of model~A3 have narrower distributions. This is because for
model~A3 the turbulent velocity, $u_t$, is much smaller than the buoyant
velocity, and thus the final positions of the bubbles are mainly
determined by buoyancy.  In the future, with more observations about the
positions of old bubbles in clusters, we could more tightly constrain
the typical velocity and scale of turbulence in cluster cores.

We note that the systematic motion of the cD galaxy containing an AGN
with respect to the ICM, $v_p$, may affect the asymmetry of bubble
positions. On the frame that moves with the cD galaxy, this can be
studied as the case of $l_t=\infty$ and $u_t=v_p$. Our results may
indicate that $v_p\gtrsim 100\:\rm km\: s^{-1}$ for the Perseus cluster,
and this could be observed by {\it SUZAKU} as the shift of metal lines
from the optical redshift of the cD galaxy.  Since the observed bubbles
are inside the potential wells of cD galaxies, the aspherical potential
of a cD galaxy could also affect of the asymmetry of bubble
positions. The former could be measured from the optical image of the
galaxy. For the Perseus cluster, the ratio of the major axis to the
minor axis in a red color is $<1.1$,\footnote{NED:
http://nedwww.ipac.caltech.edu/} which does not affect the results
above.

\section{Conclusions}

Using a simple model, we have investigated the turbulence at scales
larger than the size of bubbles observed in cool cores of galaxy
clusters. We showed that for a bubble pair, which was created by AGN jet
activities $\sim 10^8$~yrs ago, the projected angle between the two
vectors from the cluster center to the two bubbles should be $\gtrsim
90^\circ$, and the ratio of their projected distances from the cluster
center should be $\lesssim 2.5$, if the velocity and scale of turbulence
are $\sim 250\rm \: km\: s^{-1}$ and $\sim 20$~kpc, respectively. The
positions of the bubbles observed in the Perseus cluster indicate that
the velocity is $\gtrsim 100\rm\: km\: s^{-1}$ for the cluster. This
turbulence could directly be detected by {\it SUZAKU}.  \citet{fab03a}
indicated that the ICM is not turbulent at the scale of $\lesssim
10$~kpc. Detailed observations of X-ray spectra may reveal the lower
limit of the scale of turbulence \citep{ino03}, which might be
determined by the viscosity of the ICM \citep{sch04}.

\acknowledgments

I thank the anonymous referee for useful comments.  Y.~F.\ was supported
in part by a Grant-in-Aid from the Ministry of Education, Culture,
Sports, Science, and Technology of Japan (17740182).

\clearpage

\begin{deluxetable}{ccccccc}
\tablecaption{Model Parameters  \label{tab:model}}
\tablewidth{0pt}
\tablehead{
\colhead{Model} & $l_t$ & $u_t$ & $\theta_m$ & $\Gamma_m$ 
& $f_\theta(>120^\circ)$ & $f_\Gamma(<1.15)$
\\
\colhead{} & (kpc) & ($\rm km\: s^{-1}$) & (degree) & \colhead{}
& \colhead{} & \colhead{}
}
\startdata
A1       & 10   & 250  & 153 & 1.37 & 0.82 & 0.24 \\
A2       & 20   & 250  & 144 & 1.43 & 0.72 & 0.22 \\
A3       & 40   & 250  & 129 & 1.47 & 0.58 & 0.21 \\
B1       & 20   &  25  & 177 & 1.04 & 0.998 & 0.92 \\
B2       & 20   & 500  & 115 & 1.61 & 0.47 & 0.17 \\
\enddata
\end{deluxetable}

\clearpage

\clearpage

\clearpage

\end{document}